\documentclass[cits]{PoS}

\title{Transversity in exclusive electroproduction of pseudoscalar mesons}

\ShortTitle{Transversity in exclusive electroproduction of pseudoscalar mesons }

\author{\speaker{Peter Kroll}\thanks{This work is supported in part by the
    BMBF under contract 06RY258.}\\
        Universitaet Wuppertal\\
        E-mail: \email{kroll@physik.uni-wuppertal.de}}

\abstract{In this talk it is reported on an analysis of hard exclusive  
electroproduction of pseudoscalar mesons within the handbag approach. 
It is argued that recent measurements of pion electroproduction performed 
by HERMES and CLAS clearly indicate the occurence of strong contributions 
from transversely polarized photons. Within the handbag approach such 
$\gamma^{\,*}_T\to \pi$ transitions are described by the transversity GPDs 
accompanied by twist-3 pion wave functions. It is shown that this handbag 
approach leads to results on cross sections and single-spin asymmetries in 
fair agreement with experiment. The surprising result is that the $\pi^0$
cross section is dominated by $\gamma^{\,*}_T\to \pi$ transitions.
Predictions for other pseudoscalar meson channels are also discussed.}

\FullConference{Sixth International Conference on Quarks and Nuclear Physics\\
                 April 16-20, 2012\\
                 Ecole Polytechnique, Palaiseau,  Paris}

\begin{document}

\begin{figure}[b]
\includegraphics[width=0.52\textwidth] %bb=105 513 358 658,clip=true]
{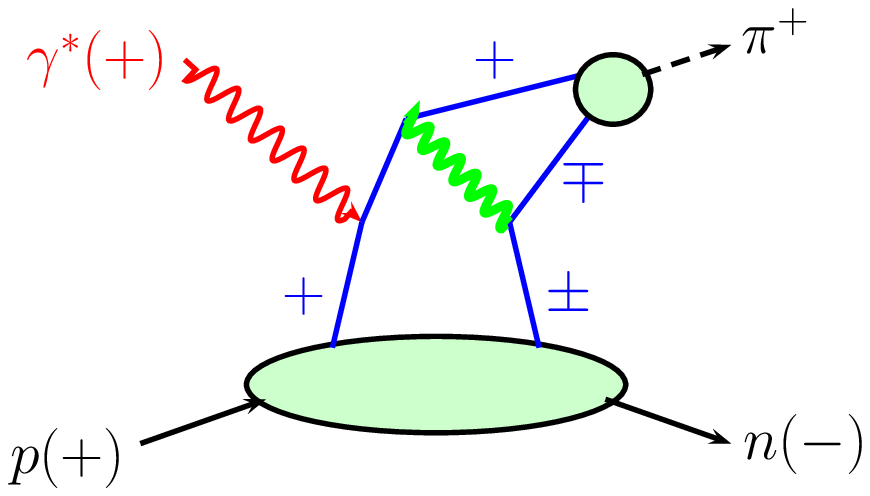}
\includegraphics[width=0.42\textwidth] %bb=25 350 532 743,clip=true]
{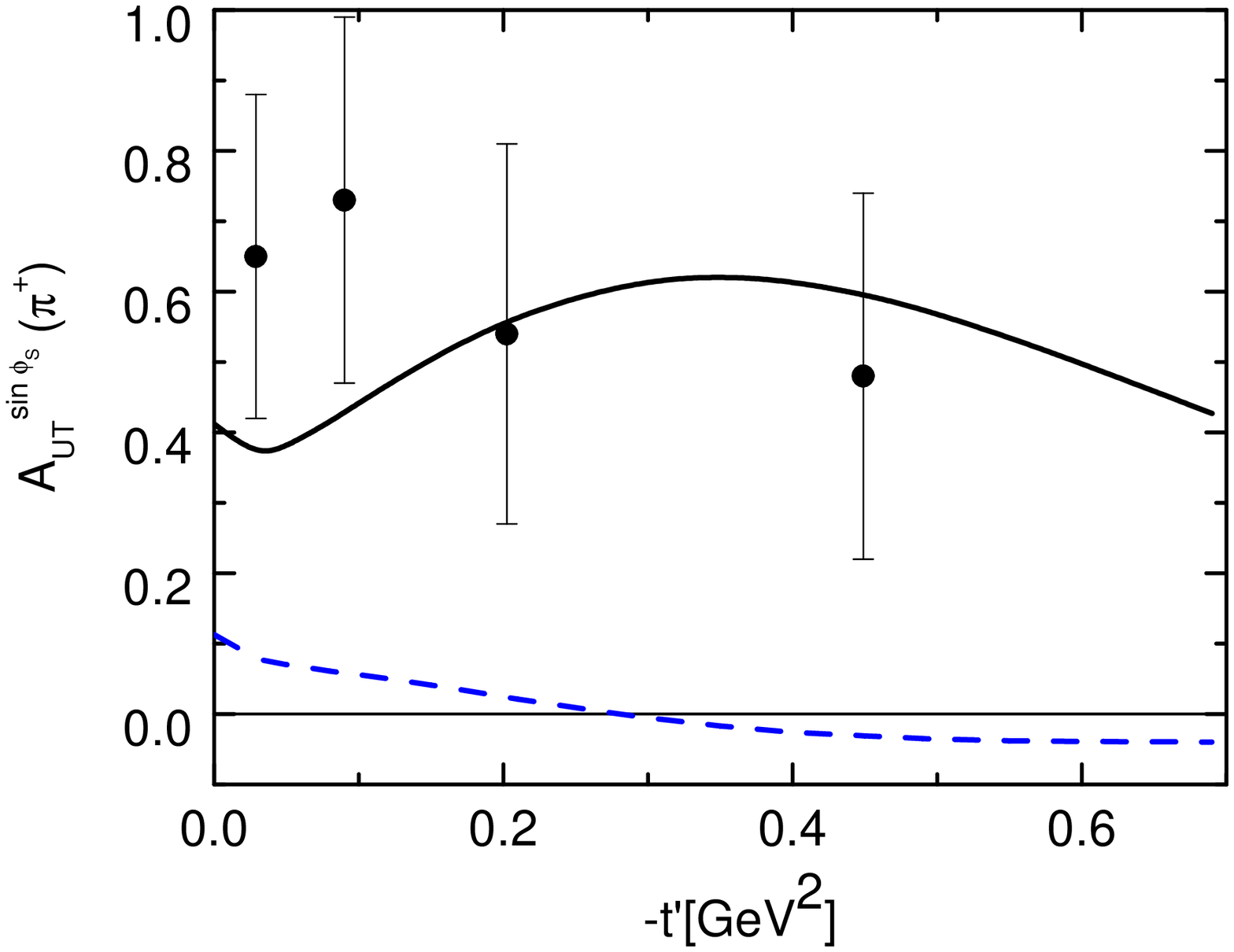}
\caption{\label{fig:1}   A typical lowest order Feynman graph for pion 
electroproduction. The signs indicate helicity labels for the handbag 
contribution to the amplitude ${\cal M}_{0-,++}$, see text.}
\caption{\label{fig:2} The $\sin{\phi_s}$ moment of the pion electroproduction
    cross section measured with a transversely polarized target at 
    $Q^2\simeq 2.45\,{\rm GeV}^2$ and $W=3.99\,{\rm GeV}$. The handbag 
    prediction~\protect\cite{GK5} is shown as a solid line. The 
    dashed line is obtained disregarding the twist-3 contribution. Data are 
    taken from Ref.\ \protect\cite{Hristova}.}
\end{figure}

In this talk it is reported upon an analysis of hard exclusive
electroproduction of pseudoscalar mesons ~\cite{GK5,GK6} within the 
framework of the handbag approach which offers a partonic 
description of meson electroproduction provided the virtuality of the 
exchanged photon, $Q^2$, and the energy, $W$, in the photon-proton center 
of mass frame are sufficiently large. The theoretical basis of the 
handbag approach is the factorization of the process amplitudes in  
hard partonic subprocesses and soft hadronic matrix elements, so-called 
generalized parton distributions (GPDs), as well as wave functions for 
the produced mesons, see Fig.\ \ref{fig:1}. In collinear approximation 
factorization has been shown~\cite{rad96,col96} to hold rigorously for 
exclusive meson electroproduction in the limit $Q^2\to\infty$. It has 
also been shown that the transitions from a longitudinally polarized 
photon to the pion, $\gamma^{\,*}_L\to \pi$, dominate at large $Q^2$. 
Transitions from transversely polarized photons to the pion are
suppressed by inverse powers of the hard scale. In Refs.\ \cite{GK5,GK6} 
a variant of the handbag approach is utilized for the interpretation of 
the data in which the subprocess amplitudes are calculated using $k_\perp$
factorization. The partons are still emitted and re-absorbed by 
the nucleon collinearly. It has been shown~\cite{GK1} that within this 
handbag approach the data on cross sections and spin density matrix 
elements for vector-meson production are well fitted for small values of 
skewness 
($\;\xi\simeq x_{Bj}/2\,{\raisebox{-4pt}{$\,\stackrel{\textstyle
                                               <}{\sim}\,$}}\, 0.1\;$).

The HERMES collaboration \cite{Hristova} has measured the $\pi^+$ 
electroproduction cross section with a transversely polarized target.
The $\sin{\phi_s}$ moment of this cross section is displayed in Fig.\ 
\ref{fig:2} ($\phi_s$ specifies the orientation of the target spin vector).
Particularly striking is the fact that the $\sin{\phi_S}$ moment         
exhibits a mild $t^\prime$-dependence and does not show any indication for a 
turnover towards zero for $t^\prime\to 0$. This behavior of 
$A_{UT}^{\sin{\phi_s}}$ at small $-t^\prime$ can only be produced by the 
interference term ${\rm Im}\big[{\cal M}_{0-,++}^*\,{\cal M}_{0+,0+}\big]$. 
Both the contributing amplitudes, one for a transversely and one for a 
longitudinally polarized photon, are helicity non-flip ones and are 
therefore not forced to vanish in the forward direction by angular 
momentum conservation. The amplitude ${\cal M}_{0-,++}$ has to be sizable 
in order to account for the HERMES data. Moreover, the amplitude 
${\cal M}_{0-,-+}$ which vanishes $\propto t^\prime$ for $t^\prime\to 0$,
cannot be large given that the $\sin{(2\phi-\phi_s)}$ moment is small 
\cite{Hristova}.

\begin{figure}[t]
\begin{center}
\includegraphics[width=0.45\textwidth]{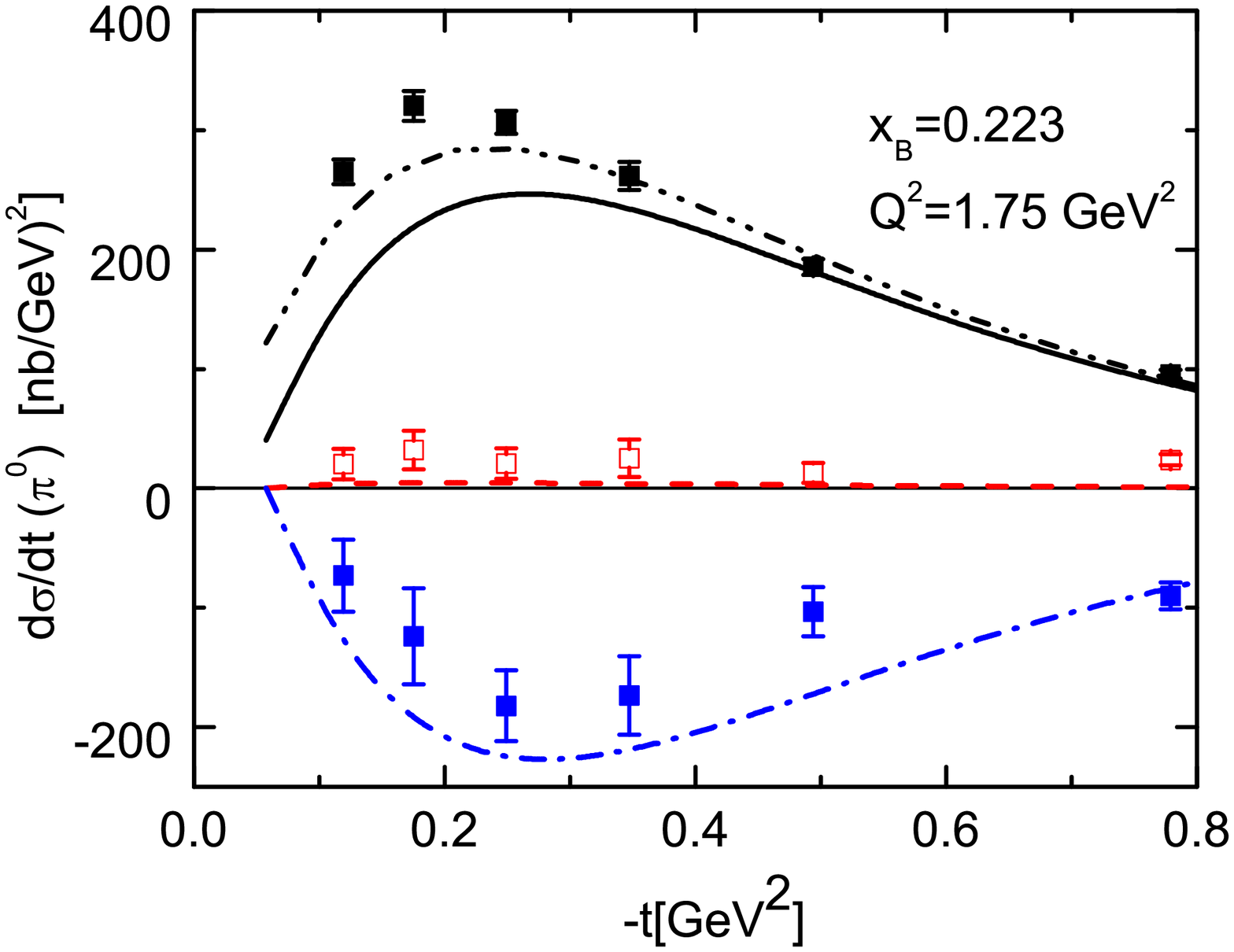}
\includegraphics[width=0.44\textwidth]{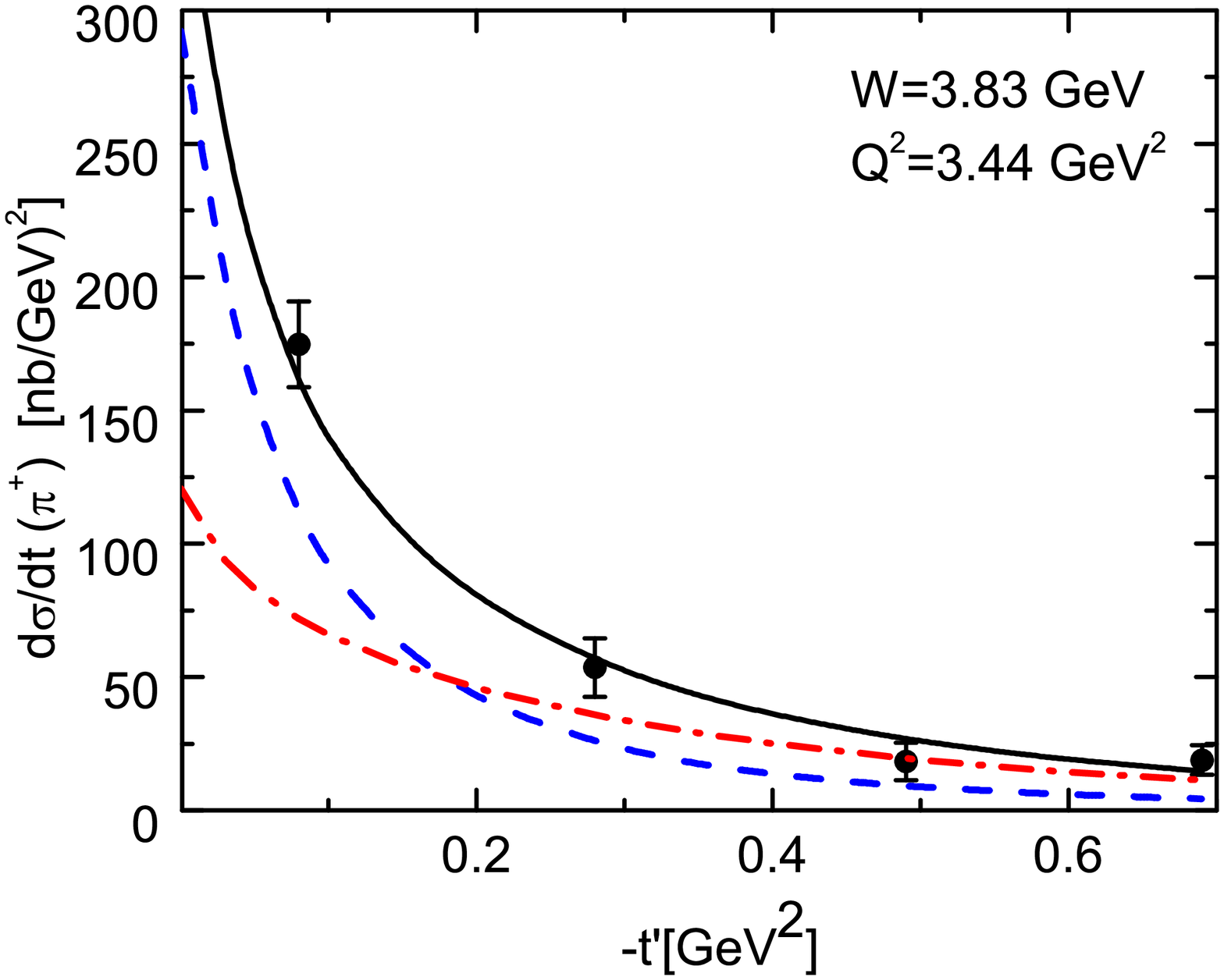}
\caption{\label{fig:3} Left: The unseparated $\pi^0$ cross section as well as
  the longitudinal-transverse (open symbols) and the transverse-transverse
  interference (solid symbols) cross section. Preliminary data are taken from
  \protect\cite{kubarowsky}. The curves represent the results obtained in 
  \protect\cite{GK6}.} 
\caption{\label{fig:4} Left: The unseparated $\pi^+$ cross section. Data taken
  from \protect\cite{HERMES07}. The solid (dashed,dash-dotted) curve
  represents the results for the unseparated (longitudinal, transverse)
  cross section \protect\cite{GK6}.}
\end{center}
\end{figure}

A second hint at large $\gamma^{\,*}_L\to \pi$ transitions comes from
the CLAS measurement of the $\pi^0$ electroproduction cross section 
\cite{kubarowsky}. As can be seen from Fig.\ \ref{fig:3} the
transverse-transverse interference cross section is negative and, in absolute
value, amounts to a substantial fraction of the unseparated cross section.
It is convenient to introduce sum and difference of the two single-flip
amplitudes (photon helicity $\mu=\pm 1$)
\begin{equation}
{\cal M}_{0+,\mu +}^{N(U)} \,=\, \frac12 \Big[ {\cal M}_{0+,\mu+}
                         \pm  {\cal M}_{0+,-\mu+}\Big] \,,
\end{equation}
which respect the symmetry relation
\begin{equation}
 {\cal M}_{0+,- +}^{N(U)} \,=\, \pm {\cal M}_{0+,+ +}^{N(U)}\,.
\end{equation}
This relation is known from one-particle exchange of either natural or
unnatural parity. If the amplitude ${\cal M}_{0-,-+}$ is neglected the
transverse and the transverse-transverse interference cross section can be
written as ($\kappa$ is a phase space factor)
\begin{eqnarray}
\frac{d\sigma_T}{dt} &=& \frac1{2\kappa}\,\Big[|{\cal M}_{0-,++}|^2
            +2 |{\cal M}^N_{0+,++}|^2 +2 |{\cal M}^U_{0+,++}|^2\Big]\,,\nonumber\\
\frac{d\sigma_{TT}}{dt} &=& -\frac1{\kappa}\,\Big[
            |{\cal M}^N_{0+,++}|^2 - |{\cal M}^U_{0+,++}|^2\Big]\,.
\end{eqnarray}
The CLAS $\pi^0$ data tell us that the amplitude ${\cal M}^N_{0+++}$
is large and ${\cal M}^U_{0+++}$ small, see Fig.\ \ref{fig:3}.

How can the amplitudes for $\gamma^*_T\to \pi$ transitions be modeled in the 
framework of the handbag approach? From Fig.\ \ref{fig:1} where the helicity 
configuration for the amplitude ${\cal M}_{0-,++}$ is indicated, it is clear 
that contributions from the usual helicity non-flip GPDs $\widetilde{H}$ and 
$\widetilde{E}$ to this amplitude do not have the properties required by the 
data on the $\sin{\phi_s}$ moment. Angular momentum conservation forces both 
the parton-nucleon vertex and the subprocess, to vanish as $\sqrt{-t^\prime}$ 
at least. There is a second set of GPDs, the helicity-flip or transversity ones 
$H_T, E_T, \ldots$ \cite{diehl01,hoodbhoy} for which the emitted and
reabsorbed partons have opposite helicities. As an inspection of Fig.\ 
\ref{fig:1} reveals the parton-nucleon vertex as well as the subprocess 
amplitude ${\cal H}_{0-,++}$ are now of helicity non-flip nature and are 
therefore not forced to vanish in the forward direction. The prize to pay is 
that quark and antiquark forming the pion have the same helicity. Therefore, 
the twist-3 pion wave function is needed instead of the familiar twist-2 one. 
This dynamical mechanism which is of twist-3 accuracy, also applies to the 
amplitudes ${\cal M}_{0+,\pm +}$. 

In Ref.\ \cite{GK5,GK6} the twist-3 pion wave function is taken from 
\cite{braun90} with the three-particle Fock component neglected. This wave 
function contains a pseudoscalar and a tensor component. The latter one
provides a contribution to the subprocess amplitude $H_{0-,++}$ which is 
proportional to $t^\prime/Q^2$ and is neglected. The contribution from the 
pseudoscalar component to $H_{0-,++}$ has the required properties. It is 
proportional to the parameter $\mu_\pi=m^2_\pi/(m_u+m_d) \simeq 2\,{\rm GeV}$ 
at the scale of $2\,{\rm GeV}$ as a consequence of the divergency of the 
axial-vector current ($m_u$ and $m_d$ are current quark masses). Although 
parametrically suppressed by $\mu_\pi/Q$ as compared to the longitudinal 
amplitudes, it is sizeable for $Q$ of the order of a few GeV. The other quark 
helicity-flip subprocess amplitude $H_{0-,-+}$ is $\propto t/Q^2$ and
therefore neglected in \cite{GK5,GK6}.

The general structure of the handbag approach for the $\gamma^*_T\to \pi$ 
amplitudes is in perfect agreement with the experimental findings discussed 
above 
\begin{eqnarray}
{\cal M}_{0-,+ +} &=& e_0 \sqrt{1-\xi^2} \int dx\; H_{0-+ +}\; H_T +
  {\cal O}(\xi^2)\,  \nonumber\\
{\cal M}_{0+,\pm +} &=& -e_0 \frac{\sqrt{-t^\prime}}{4m} \int dx\;
  H_{0-+ +}\; \overline{E}_T + {\cal O}(\xi^2)\,,
\end{eqnarray}
where $\overline{E}_T\equiv 2\widetilde{H}_T + E_T$ and ${\cal M}_{0+,\pm +}$ 
behaves like a natural parity exchange; the unnatural part is 
${\cal O}(\xi)$ and neglected as well as the double-flip amplitude 
${\cal M}_{0-,-+}$ which behaves $\propto t^\prime$.       

In order to make predictions also the GPDs are needed. In \cite{GK5,GK6}
they are constructed with the help of the double distribution ansatz
\cite{rad98} consisting of the product of a zero-skewness GPD and an
appropriate weight function which generates the skewness dependence. The
zero-skewness GPDs are parameterized as their forward limits multiplied by a
Regge-like $t$ dependence, $\exp{[t(b_i-\alpha_i^\prime\ln{x})]}$. In the case
of $\widetilde{H}$ the forward limit is given by the polarized parton 
distributions. The GPD $H_T$ is constrained by the transversity PDF 
$\delta(x)$ for which the results of an analysis of the asymmetries in 
semi-inclusive electroproduction are taken \cite{anselmino}. The lowest
moments of this variant of $H_T$ are smaller by about a factor of 2 than 
lattice QCD results \cite{haegler}. Therefore, an alternative variant of 
$H_T$ is also considered which is normalized to the lattice results 
\cite{haegler}. The second transversity GPD $\overline{E}_T$ is parameterized 
in the same spirit as the others and normalized to the lattice results as well
because other information on it is lacking at present. It is important to
stress that $\overline{E}_T$ has the same sign and almost the same size for
$u$ and $d$ quarks in which aspect it differs from $H_T$. The remaining 
parameters of the GPDs are fitted to the only available small-skewness data, 
namely the $\pi^+$ electroproduction data from HERMES
\cite{Hristova,HERMES07}. With regard to the uncertainties in the
parameterization of the GPDs the predictions for pseudoscalar meson 
electroproduction given in \cite{GK6}, with the exception of $\pi^+$ at small 
skewness, are to be considered as estimates of trends and magnitudes. 
  
\begin{figure}[t]
\begin{center}
\includegraphics[width=0.42\textwidth]{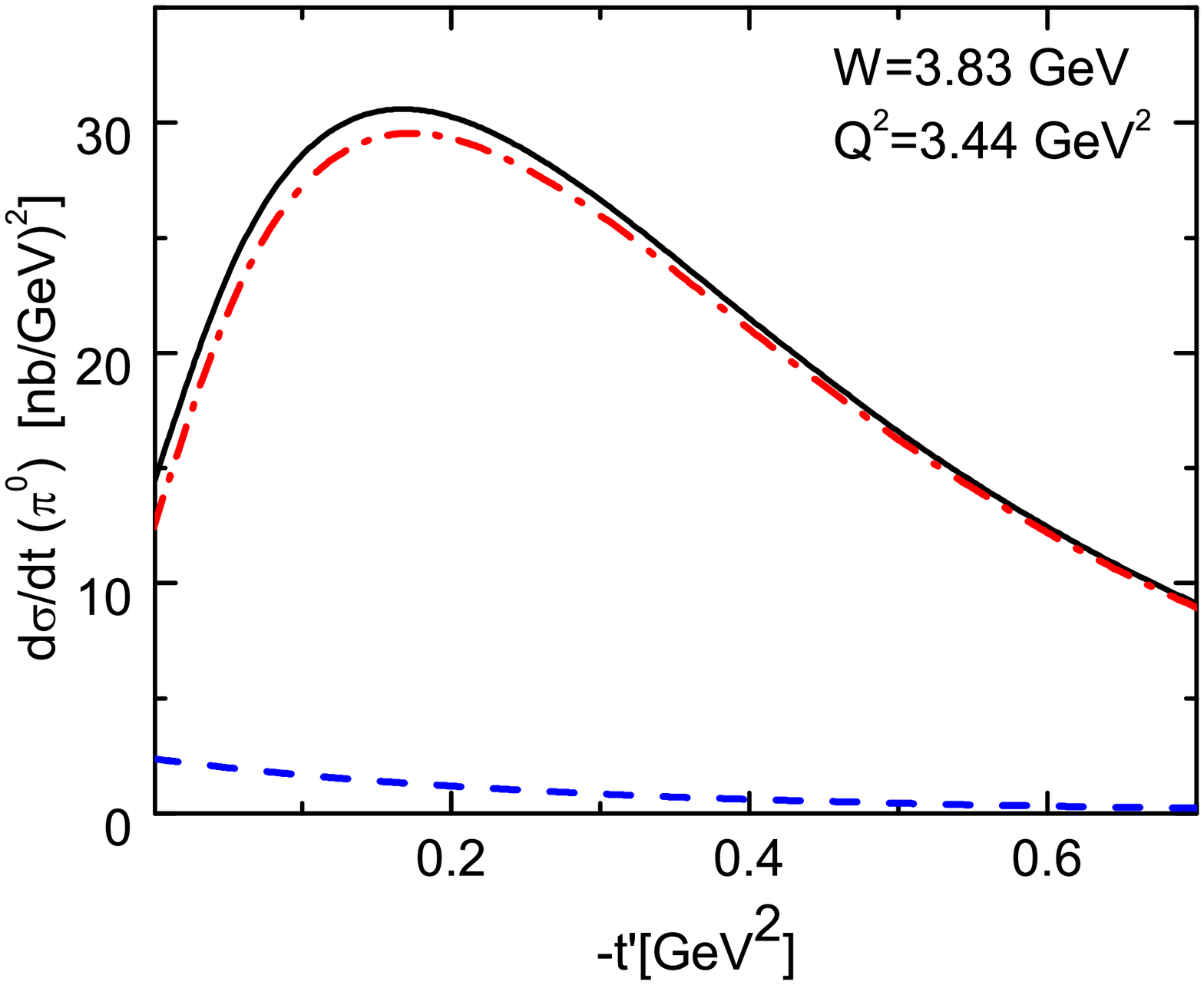}
\includegraphics[width=0.43\textwidth]{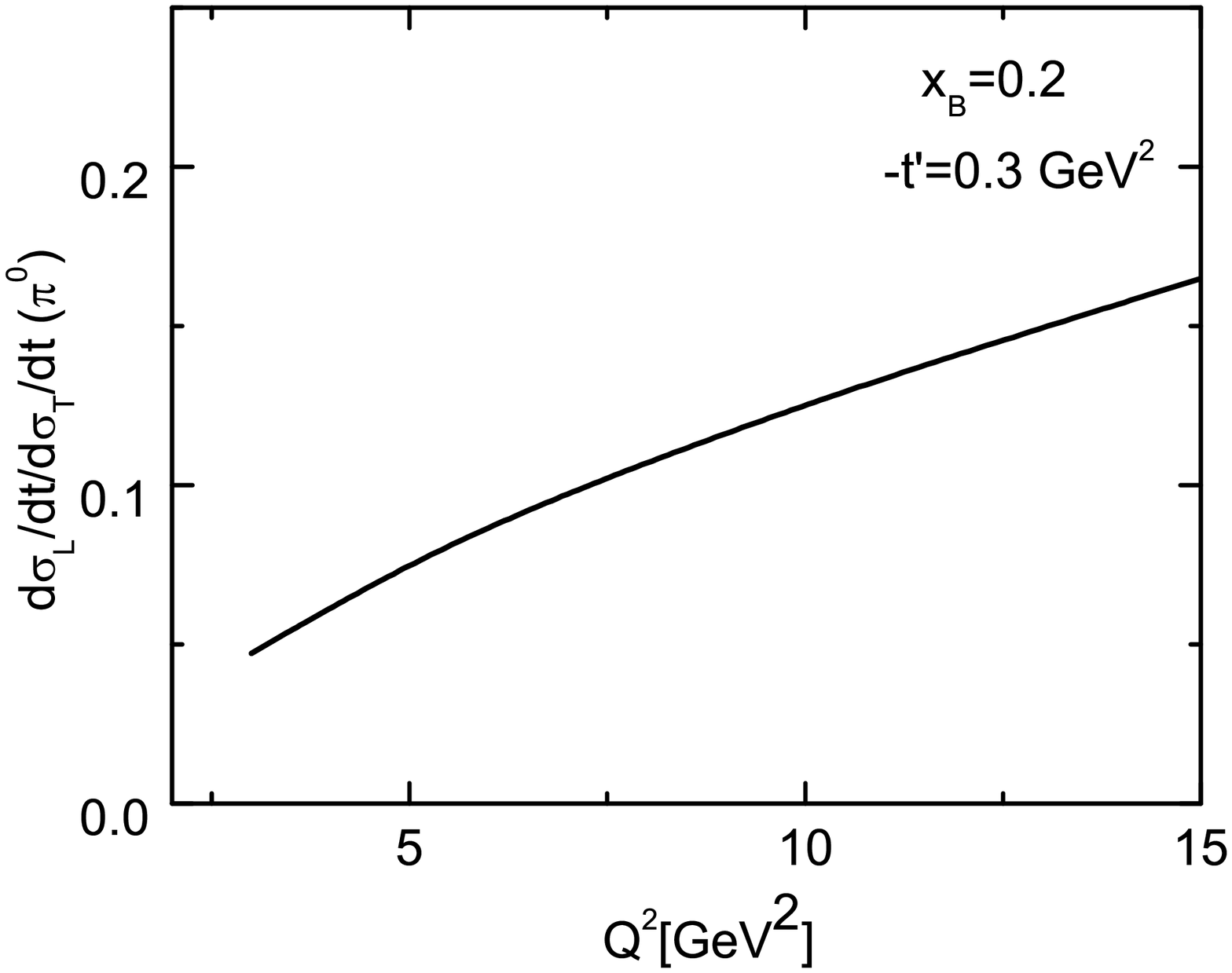}
\caption{\label{fig:5} Left: As Fig. \protect\ref{fig:4} but for $\pi^0$
electroproduction. The alternative parameterization of $H_T$ is used.}
\caption{\label{fig:6} Right: The ratio of the longitudinal and transverse
cross section for $\pi^0$ electroproduction.}
\end{center}
\end{figure}

\begin{figure}[t]
\begin{center}
\includegraphics[width=0.43\textwidth]{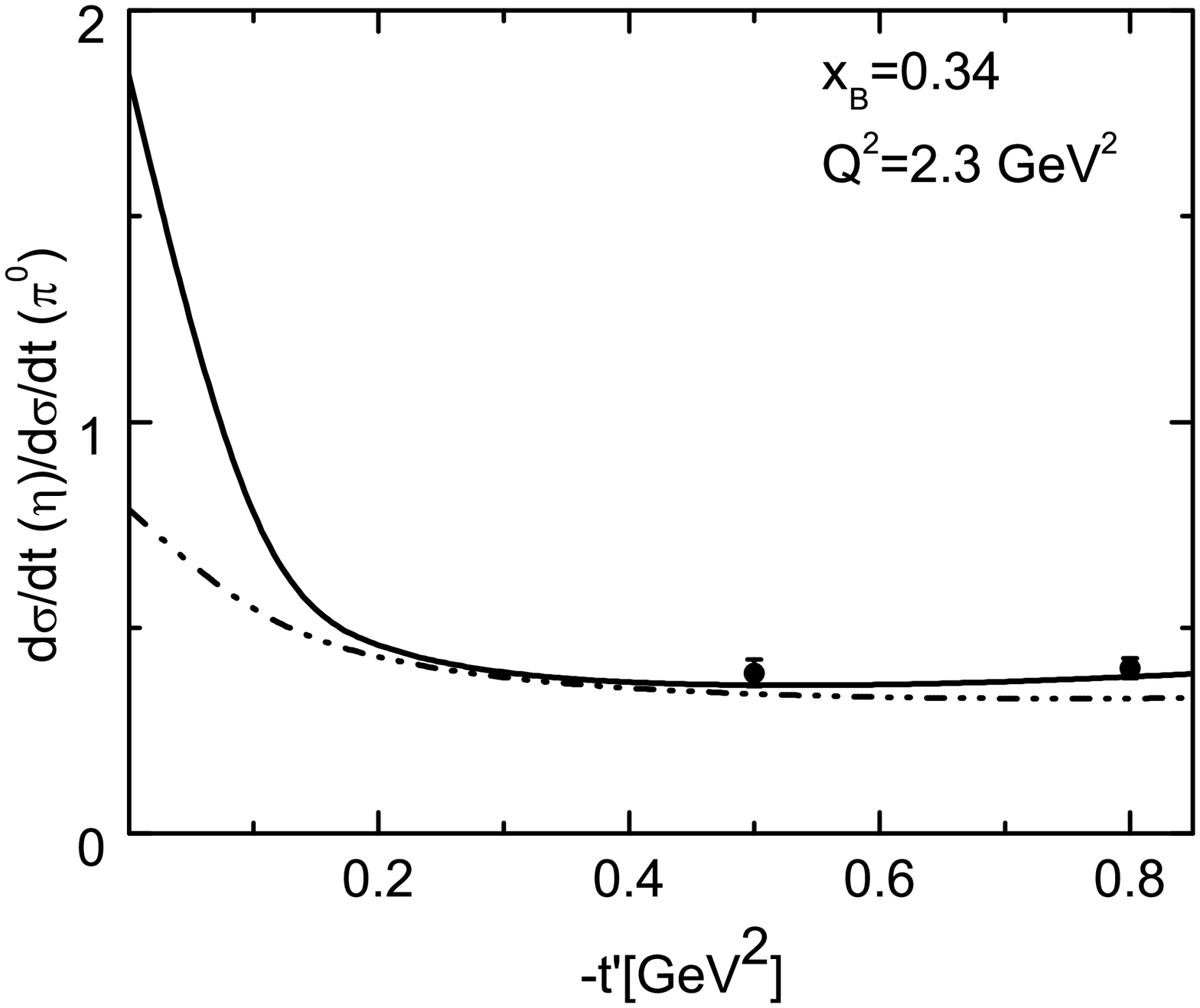}
\includegraphics[width=0.435\textwidth]{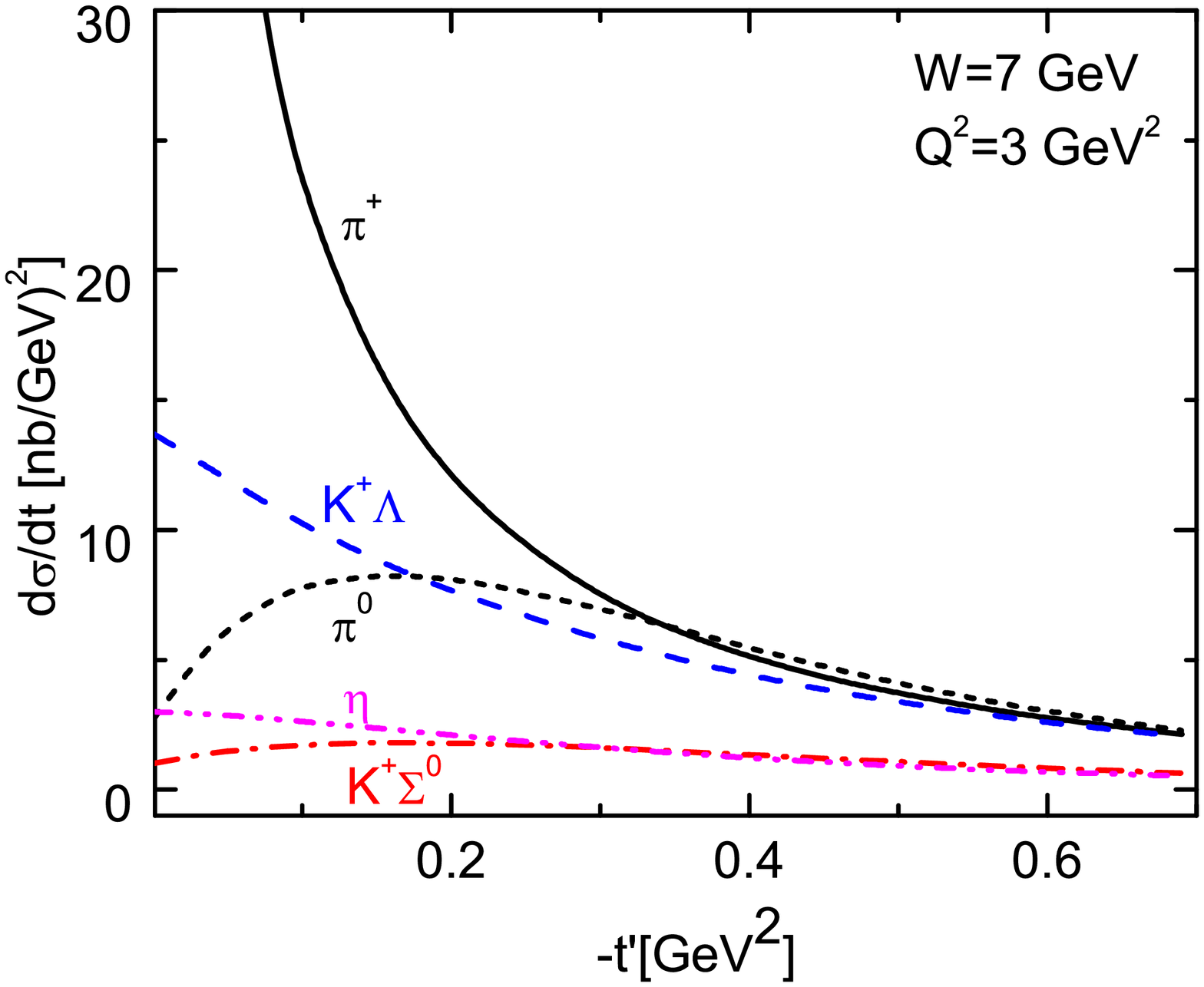}
\caption{ \label{fig:7} Left: The ratio of the $\eta$ and $\pi^0$ cross
  sections versus $-t^\prime$. The predictions given in \cite{GK6} are shown
 as a solid line. The dash-dot-dotted line is the result obtained with the
 alternative variant of $H_T$. The preliminary data are taken from 
\protect\cite{kubarowsky}.}
\caption{\label{fig:8} Right:The cross sections for various pseudoscalar meson
  channels.}   
\end{center}
\end{figure}

A few of the results obtained in \cite{GK5,GK6} are shown in Figs.\ \ref{fig:2}
-- \ref{fig:8}. As can be seen from Figs.\ \ref{fig:2} and \ref{fig:4}
the transverse target asymmetries~\cite{Hristova} as well as the cross section
\cite{HERMES07} for $\pi^+$ electroproduction are nicely fitted. The prominent
role of the twist-3 mechanism for understanding the behavior of the
$\sin{\phi_s}$ moment is obvious from the two curves shown in Fig.\
\ref{fig:2}. While the $\pi^+$ cross section obtains substantial contributions 
from both longitudinally polarized photons (at small $-t^\prime$) and
transverse ones (at large $-t^\prime$) is the $\pi^0$ cross section strongly
dominated by the $\gamma^*\to \pi^0$ transitions, see Figs.\
\ref{fig:3}--\ref{fig:5}. The strong dip of the forward cross section signals 
the dominance of the single helicity-flip amplitudes ${\cal M}_{0+\pm +}$,
i.e. of contributions from $\overline{E}_T$. Although the $\gamma^*\to \pi$ 
transitions are suppressed by $\mu_\pi/Q$ as compared to the asymptotically 
dominant contributions from longitudinally polarized photons the ratio 
$\sigma_L/\sigma_T$ is very small for $\pi^0$ production at small $Q^2$  but
it increases with $Q^2$, see Fig.\ \ref{fig:6}. The longitudinal cross section 
takes the lead only for very large values of $Q^2$.

In Fig.\ \ref{fig:7} the ratio of the $\eta$ and $\pi^0$ cross section is
shown. Except in the proximity of the forward direction where the contributions 
from $H_T$ dominate, the ratio is small and in good agreement with preliminary 
CLAS data \cite{kubarowsky}. The smallness of the ratio is a consequence of the 
properties of the dominant GPD $\overline{E}_T$, namely the same signs and
about the same size of $\overline{E}^u_T$ and $\overline{E}^d_T$. Finally,
in Fig.\ \ref{fig:8} predictions for the cross sections of various
pseudoscalar meson channels are shown for typical COMPASS kinematics.
Except of the case of the $\pi^+$ all channels are dominated by  
$\gamma^*_T\to$ meson transitions although the degree of dominance differs.

In summary, there is strong evidence for transversity in hard exclusive 
electroproduction of pions. The most striking effects are seen in the
experimental data on the $\pi^+$ target asymmetry $A_{UT}^{\sin \phi_s}$ and
on the $\pi^0$ cross section. The interpretation of these effects requires a 
large helicity non-flip amplitude ${\cal M}_{0-,++}$ and the flip amplitudes 
${\cal M}_{0+++}\simeq {\cal M}_{0+-+}$. Within the handbag approach these
amplitudes are generated by the helicity-flip or transversity GPDs in 
combination with a twist-3 pion wave function~\cite{GK5,GK6}. The GPDs are 
constructed from double distributions. They are fitted to the HERMES 
small-skewness data on $\pi^+$ and are therefore optimized for that region. At 
larger values of skewness the parameterizations of the GPDs are perhaps 
to simple and may require improvements. It is also important to 
realize that the GPDs are probed by the HERMES data only for $x$ less than 
about 0.6. This does not mean that one cannot compare with experimental data 
from Jefferson Lab., e.g.\ \cite{kubarowsky} but one cannot expect that all 
details of the data will be correctly reproduced. However, as is shown, the 
trends and magnitudes of the CLAS data are reasonably well explained. Further 
studies of transversity in exclusive reactions are certainly demanded. Good 
data on $\pi^0$ electroproduction from the upgraded Jlab or from the COMPASS 
experiment would be welcome. They would not only allow for further tests of 
the twist-3 mechanism but also provide the opportunity to verify the 
parameterizations of the GPDs $\widetilde{H}$ and $\widetilde{E}$ as used in 
Ref.\ \cite{GK5,GK6}.


\begin{thebibliography}{99}
\bibitem{GK5} S.V.~Goloskokov and P.~Kroll,
  %``An attempt to understand exclusive pi+ electroproduction,''
  \emph{Eur.\ Phys.\ J.} {\textbf C65} (2010)~137,
  [arXiv:0906.0460].
  %%CITATION = ARXIV:0906.0460;%%

\bibitem{GK6} S.~V.~Goloskokov and P.~Kroll,
  %``Transversity in hard exclusive electroproduction of pseudoscalar mesons,''
  Eur.\ Phys.\ J.\ A {\bf 47}, 112 (2011)
  [arXiv:1106.4897 [hep-ph]].
  %%CITATION = ARXIV:1106.4897;%%

\bibitem{rad96}A.~V.~Radyushkin,
  %``Asymmetric gluon distributions and hard diffractive electroproduction,''
  \emph{Phys.\ Lett.}  {\textbf B385} (1996)~333, [hep-ph/9605431].
  %%CITATION = PHLTA,B385,333;%%

\bibitem{col96} J.C.\ Collins, L.\ Frankfurt and M.\ Strikman, 
  %``Factorization for hard exclusive electroproduction of mesons in QCD,''
  \emph{Phys.\ Rev.} {\textbf D56}  (1997)~2982, [hep-ph/9611433].
    %%CITATION = HEP-PH 9611433;%%

\bibitem{GK1} S.~V.~Goloskokov and P.~Kroll,
    %``Vector meson electroproduction at small Bjorken-x and generalized parton
    %distributions,''
   \emph{Eur.\ Phys.\ J.} {\textbf C42} (2005)~281;
%   {\it ibid.}  {\textbf C50} (2007)~829;
   {\it ibid.}  {\textbf C53} (2008)~367.

\bibitem{Hristova} A.~Airapetian {\it et al.}  [HERMES Collaboration],
  %``Single-spin azimuthal asymmetry in exclusive electroproduction of pi+
  %mesons on transversely polarized protons,''
\emph{Phys.\ Lett.} {\bf B682}, 345 (2010),
  [arXiv:0907.2596].
  %%CITATION = ARXIV:0907.2596;%%

\bibitem{kubarowsky}V. Kubarovsky {\it et al}, 
  % "Deeply Virtual Pseudoscalar Meson Production with CLAS"
  Proceedings of the 4th Workshop " Exclusive reactions at High Momentum
  Transfer", Newport News, VA USA, 18-21 May 2010


\bibitem{diehl01} M.~Diehl,
  \emph{Eur.\ Phys.\ J.} {\textbf C19} (2001) 485,
  [hep-ph/0101335].
  %%CITATION = ARXIV:0906.0460;%%

\bibitem{hoodbhoy} P.~Hoodbhoy and X.~Ji,
  \emph{Phys.\ Rev.} {\textbf D58} (1998)~054006, [hep-ph/9801369].

\bibitem{braun90} V.~M.~Braun and I.~E.~Halperin,
  %``Conformal Invariance And Pion Wave Functions Of Nonleading Twist,''
  \emph{Z.\ Phys.} {\textbf C48} (1990)~239.
  [\emph{Sov.\ J.\ Nucl.\ Phys.}  {\textbf 52} (1990\ YAFIA,52,199-213.1990)~126].
  %%CITATION = ZEPYA,C48,239;%%


\bibitem{rad98} A.~V.~Radyushkin,
    %``Symmetries and structure of skewed and double distributions,''
    \emph{Phys.\ Lett.} {\textbf B449} (1999)~81, [hep-ph/9810466].
    %%CITATION = HEP-PH 9810466;%%

\bibitem{anselmino} M.~Anselmino {\it et al.},
    %``Transversity and Collins functions from SIDIS and e+ e- data,''
    \emph{Phys.\ Rev.} {\textbf D75} (2007)~054032, [hep-ph/0701006]. 
    %%CITATION = PHRVA,D75,054032;%%

\bibitem{haegler} Ph.~Hagler {\it et al.}  [LHPC Collaborations],
  %``Nucleon Generalized Parton Distributions from Full Lattice QCD,''
  Phys.\ Rev.\  D {\bf 77}, 094502 (2008)
  [arXiv:0705.4295]. %[hep-lat]].
  %%CITATION = PHRVA,D77,094502;%%

\bibitem{HERMES07} A.~Airapetian {\it et al.}  [HERMES Collaboration],
    %``Cross sections for hard exclusive electroproduction of pi+ mesons on a
    %hydrogen target,''
    \emph{Phys.\ Lett.}  {\textbf B659} (2008)~486, [arXiv:0707.0222].
    %%CITATION = PHLTA,B659,486;%%

\end{thebibliography}
\end{document}